\documentclass[aip,rsi,reprint]{revtex4-1}
\usepackage{graphicx}
\usepackage{subfigure}
\usepackage{mathtools}
\usepackage{empheq}
\usepackage{color}
\begin{document}

% Use the \preprint command to place your local institutional report number 
% on the title page in preprint mode.
% Multiple \preprint commands are allowed.
%\preprint{}

\title{An analytical model for the detection of levitated nanoparticles in optomechanics} %Title of paper

\author{A. T. M. Anishur Rahman}
\email{a.rahman@ucl.ac.uk}
\affiliation{Department of Physics and Astronomy, University College London, Gower Street,  WC1E 6BT, UK}
\affiliation{Department of Physics, University of Warwick, Gibbet Hill Road, CV4 7AL, UK}
\author{A. C. Frangeskou}
\affiliation{Department of Physics, University of Warwick, Gibbet Hill Road, CV4 7AL, UK}
\author{P. F. Barker}
\affiliation{Department of Physics and Astronomy, University College London, Gower Street,  WC1E 6BT, UK}
\author{G. W. Morley}
%%\email{gavin.morley@warwick.ac.uk}
\affiliation{Department of Physics, University of Warwick, Gibbet Hill Road, CV4 7AL, UK}

% repeat the \author .. \affiliation  etc. as needed
% \email, \thanks, \homepage, \altaffiliation all apply to the current author.
% Explanatory text should go in the []'s, 
% actual e-mail address or url should go in the {}'s for \email and \homepage.
% Please use the appropriate macro for the type of information

% \affiliation command applies to all authors since the last \affiliation command. 
% The \affiliation command should follow the other information.

\begin{abstract}
Interferometric position detection of levitated particles is crucial for the centre-of-mass (CM) motion cooling and manipulation of levitated particles. In combination with balanced detection and feedback cooling, this system has provided picometer scale position sensitivity, zeptonewton force detection, and sub-millikelvin CM temperatures. In this article, we develop an analytical model of this detection system and compare its performance with experimental results allowing us to explain the presence of spurious frequencies in the spectra.
% insert abstract here
\end{abstract}

\pacs{}% insert suggested PACS numbers in braces on next line

\maketitle %\maketitle must follow title, authors, abstract and \pacs
%\title{Balanced detector}
%\maketitle %% null function with osajnl.sty

%\begin{eqnarray}
%g(x,y)&\approx& h_0F\bigl(\frac{x}{\lambda d},\frac{y}{\lambda d}\bigr)
%\end{eqnarray}. 

\section{Introduction}

In the past interferometric position detection systems have been used in optomechanics for the detection of zeptonewton scale forces\cite{GieselerNatPhys2013,GambhirPRA2015,GambhirPRA2016}, the demonstration of sub-Kelvin centre-of-mass temperatures\cite{LiNatPhys2011,Gieseler2012,Jain2016}, the measurement of Brownian motions\cite{LiScience2010}, and the manipulation of levitated particles\cite{GieselerNatNano2014,GieselerPRL2014,Neukirch2015,RahmanSciRep2016,Vovrosh2017}. Furthermore, this system has provided pm/$\sqrt{Hz}$ position sensitivity\cite{LiNatPhys2011}. In these schemes a reference beam and the scattered light from a levitated particle interfere on a photodiode. This interference produces a signal which is directly related to the instantaneous position of the oscillator. After Fourier transformation, oscillation frequencies ($\omega_x$, $\omega_y$, and $\omega_z$) along the three axes can be retrieved from the position signals. Subsequently, these frequencies are used for parametric feedback cooling to actively control the motion of a levitated particle\cite{,Gieseler2012,GieselerNatPhys2013,GieselerNatNano2014,GieselerPRL2014,Neukirch2015,GambhirPRA2015,WanPRA2016,RahmanSciRep2016,GambhirPRA2016,Vovrosh2017,RahmanNatPhot2017}. As with other interferometric schemes, this system is well-known for its high precision and resilience to noise. In optomechanical set-ups this is further enhanced by a balanced detection system. A balanced detector consists of two matched photodiodes which help to reduce common mode noise and other unwanted signals. Here, we develop a model of this interferometric scheme and present experimental evidence to justify its validity. We find that the predictions of our model match closely with the experimental results. We also show that due to the configuration of the balanced detector, it detects frequency along the desired axis as well as frequencies from the remaining two axes and the frequencies resulting from the various linear combinations of $\omega_x$, $\omega_y$, and $\omega_z$. Finally, we discuss the possible side effects of these spurious frequencies on the performance of parametric feedback cooling.

\section{Interferometric detection scheme}

\begin{figure}[h]
\centering
\includegraphics[width=8.5cm]{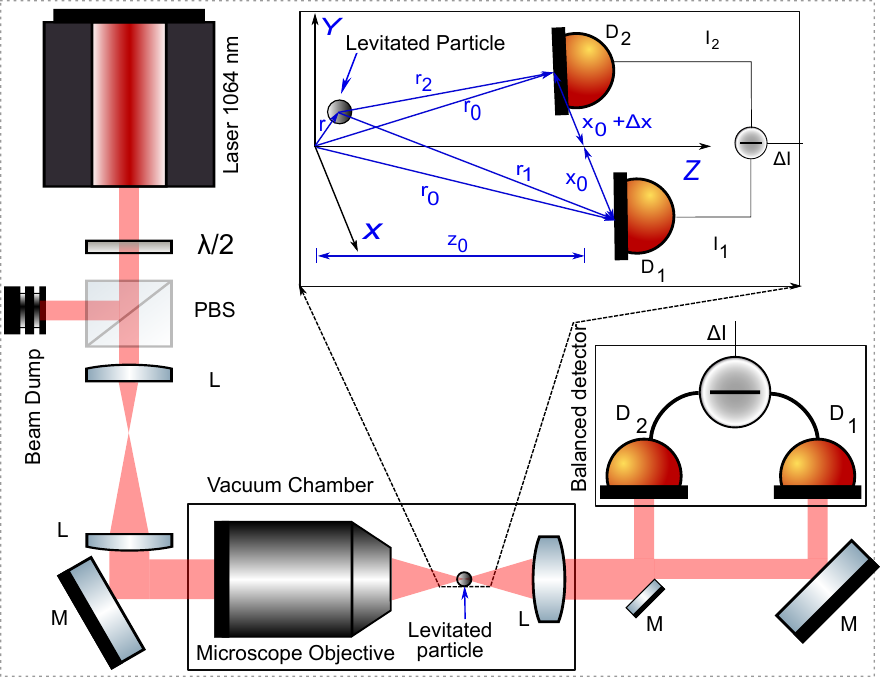}
\caption{A schematic of our tweezers based optomechanical system along with the interferometric detection system along the $x-$axis. The origin of the coordinate system is the centre of the trap. The inset shows the levitated particle along with the detection system in the co-ordinate system. Note that $z_0$ actually signifies the distance between the levitated particle and the lens after it. Since the light is collimated after the lens, the distance between the lens and the diodes, shown for the sake of visualization in the main schematic, is not important. Further, different symbols correspond to L-lens, M-mirror, PBS- polarizing beam splitter, $\lambda/2$ - halfwave plate and D-diodes.}
\label{fig1}
\end{figure}

%%$|\mathbf{r_1^0}|=|\mathbf{r_2^0}|=r_0$
Figure \ref{fig1} shows a schematic of a tweezer based optomechanical experiment in which a high numerical aperture microscope objective forms the trap by tightly focussing a laser beam into a diffraction limited spot. The trap is normally placed inside a vacuum chamber. Once a desired particle is trapped, the chamber is evacuated and the position of the particle is monitored using the interferometric detection system. Let us assume that $\textbf{r}=x\hat{x}+y\hat{y}+z\hat{z}$ is the instantaneous position vector of the levitated particle from the centre of the trap, where $x=A_x\sin{(\omega_xt+\phi_x)}$, $y=A_y\sin{(\omega_yt+\phi_y)}$ and $z=A_z\sin{(\omega_zt+\phi_z)}$ are the instantaneous distances of the particle along the $x$, $y$ and $z$ axes. The angular trap frequencies are $\omega_x$, $\omega_y$ and $\omega_z$, and $A_x$, $A_y$ and $A_z$ are the respective amplitudes of oscillations along the three axes. Likewise, $\phi_x$, $\phi_y$ and $\phi_z$ are the phases along the three axes. In order to detect and manipulate the position of the levitated particle, balanced photo-detectors are placed along the various axes. As an example, in Fig. \ref{fig1}, we show one detector placed along the $x-$ axis. This enables us to detect the trap frequency along the $x-$ axis. From the geometry of the problem, the position vectors of the two photodiodes ($D_1$ and $D_2$ in Fig. \ref{fig1}) from the centre of the trap are $\mathbf{r_1^0}=x_0\hat{x}+y_0\hat{y}+z_0\hat{z}$ and $\mathbf{r_2^0}=-(x_0+\Delta x)\hat{x}+y_0\hat{y}+z_0\hat{z}$, where $x_0$, $y_0$, and $z_0$ are the distances of the two photodiodes from the centre of the trap. $\Delta x$ is the position mismatch between the two photodiodes along the $x-$ axis. This mismatch initiates an imbalance in the detector (see below for details). The distances of a levitated particle from the two photodiodes of the balanced detector can be written as $r_1=|-\mathbf{r}+\mathbf{r_1^0}|=\sqrt{r_0^2+r^2-2x_0x-2y_0y-2z_0z}$, and $r_2=|-\mathbf{r}+\mathbf{r_2^0}|=\sqrt{r_0^2+r^2+2x_0x-2y_0y-2z_0z+\Delta x(\Delta x+2x+2x_0)}$, where $r_0=\sqrt{x_0^2+y_0^2+z_0^2}$.

Let us also assume that at the focus of the trap the $y-$polarized electric field can be expressed\cite{SalehTeich,NovotnyHecht2012} as $\mathbf{E}_l=\frac{E_0w_0}{w(z)}\exp{\Bigl[-\frac{x^2+y^2}{w(z)^2}\Bigr]}\exp{\Bigr[i\omega t-ikz-i\frac{k(x^2+y^2)}{2R(z)}+i\zeta(z)\Bigr]}\hat{y}$, where $k=2\pi/\lambda$, $\omega=2\pi c/\lambda$, and $\lambda$ and $c$ are the trapping laser wavelength and speed in free space, respectively. $w(z)=w_0\sqrt{1+z^2/z_r^2}$, $R(z)=z(1+z_r^2/z^2)$, $\zeta=\tan^{-1}{z/z_r}$ and $w_0=\sqrt{\lambda z_r/\pi}$, where $z_r$ is the Rayleigh range. $E_0$ can be expressed as $\sqrt{2I_0/\epsilon_0 c}$, where $I_0$ is the intensity of a Gaussian trapping laser beam at the focus and $\epsilon_0$ is the dielectric constant of free space. The electric field induces a dipole moment in the trapped particle. This leads to a surface charge density if the polarization is uniform throughout the trapped bead or a volume charge density otherwise \cite{FeynmanVol_1}.

Once a charge is induced inside a particle, it starts to oscillate in the oscillating trapping field, and an oscillating charge radiates/scatters light. The scattered field from a Rayleigh spherical particle ($a<<\lambda$) that a photodiode receives can be expressed as\cite{BohrenHuffman}

\tiny
%\begin{widetext}
\begin{eqnarray}
%\nonumber
\mathbf{E_{s_1}}&\approx&-\frac{Ak^2}{4\pi r_1^3}\Bigl(x_0^2+z_0^2+x^2+z^2-2x_0x-2z_0z\Bigr)\\
\nonumber
&&\times E_0\exp{\{i(\omega t-kz_0(1+\frac{x_0^2+y_0^2}{2z_0^2}+\frac{r^2-2x_0x-2y_0y-2z_0z}{2z_0^2}))\}}\hat{y}\\
\nonumber
\mathbf{E_{s_2}}&\approx&-\frac{Ak^2}{4\pi r_2^3}\Bigl(x_0^2+z_0^2+x^2+z^2+2x_0x+2\Delta xx_0-2z_0z\Bigr) \\
\nonumber
&&\times E_0\exp{\{i(\omega t-kz_0(1+\frac{x_0^2+y_0^2}{2z_0^2}+\frac{r^2+2x_0x-2y_0y-2z_0z}{2z_0^2}+\frac{\Delta x x_0}{z_0^2}))\}} \hat{y}
\end{eqnarray}
%\end{widetext}

\normalsize
where $A=4\pi a^3\epsilon_0(\epsilon-1)/(\epsilon+2)$, $\epsilon$ and $a$ are the polarizablity, dielectric constant and radius of the levitated particle, respectively. We have also assumed that the electric field ($E_0$) remains constant over the distance a levitated particle traverses inside the trap. This is valid when the amplitude of oscillation of a levitated particle is small compared to the beam waist $w_0=\lambda/(\pi NA)$, where NA is the numerical aperture of the trapping lens and $\lambda$ is the trapping laser wavelength.

In addition to the scattered light from the levitated particle, each photodiode also receives directly transmitted laser light from the trapping beam. In the far-field where $z_0 >> z_r$ and $(x_0^2+y_0^2) << z_0^2$, the directly transmitted beam unperturbed by the levitated particle can be expressed as\cite{SalehTeich} (see Appendix Eqs \ref{eqn100} and \ref{eqn101})

\tiny
%\begin{widetext}
\begin{eqnarray}
\nonumber
\mathbf{E_{T_1}}&\approx&\frac{z_rE_0}{z_0}\exp{\Bigl[i\omega t-ikz_0(1+\frac{x_0^2+y_0^2}{2z_0^2})+\pi/2\Bigr]}\hat{y}\\
%\nonumber
\mathbf{E_{T_2}}&\approx &\frac{z_rE_0}{z_0}\exp{\Bigl[i\omega t-ikz_0(1+\frac{x_0^2+y_0^2}{2z_0^2})+\pi/2\Bigr]}\exp{(-i\frac{\Delta x x_0k}{z_0})}\hat{y},
\end{eqnarray}
%\end{widetext}
\normalsize
where $\pi/2$ is the Gouy phase shift.

Considering the scattered field, and the field due to the directly transmitted light together, the difference in intensity $\Delta I= I_{D_2}-I_{D_1}$ that a balanced detector produces(see appendix Eqs \ref{eqn106}-\ref{eqn108} for derivations) can be written as 

\tiny
\begin{widetext}
\begin{eqnarray}
\nonumber
\Delta I&\approx&\overbrace{\frac{x_0xA^2k^4 }{2\pi^2 z_0^6}\Bigl(x_0^2+z_0^2+x^2+z^2-2z_0z\Bigr)I_0}^{Scattering}-\overbrace{\frac{Ak^2}{\pi z_0^3}\Bigl(x_0^2+z_0^2+x^2+z^2-2z_0z\Bigr)\cos{\Bigl(k\frac{r^2-2y_0y-2z_0z}{2z_0}\Bigr)}\sin{\Bigl(\frac{x_0xk}{z_0}\Bigr)}\frac{z_r}{z_0}I_0}^{Interference}\\
%\nonumber
&&-\overbrace{\frac{2x_0xAk^2 }{\pi z_0^3}\sin{\Bigl(k\frac{r^2-2y_0y-2z_0z}{2z_0}\Bigr)}\cos{\Bigl(\frac{x_0xk}{z_0}\Bigr)}\frac{z_r\epsilon_0 c E_0^2}{2z_0}}^{Interference}-\overbrace{\frac{\Delta x x_0Ak^3}{2\pi z_0^4}\Bigl(x_0^2+z_0^2+x^2+z^2-2z_0z+2x_0x\Bigr)\cos{\Bigl(k(\frac{r^2-2y_0y-2z_0z}{2z_0}+\frac{x_0x}{z_0})\Bigr)}\frac{z_r}{z_0}I_0}^{Imbalance},
%\nonumber
%%&&
%\overbrace{\frac{x_0xA^2k^4 }{2\pi^2 z_0^6}\Bigl(x_0^2+z_0^2+x^2+z^2-2z_0z\Bigr)I_0}^{Scattering}+\overbrace{\frac{z_rAk^2}{\pi z_0^4}\Bigl(x_0^2+z_0^2+x^2+z^2-2z_0z -2x_0x\Bigr)\cos{\Bigl(k\frac{r^2-2y_0y-2z_0z}{2z_0}\Bigr)}\sin{\Bigl(\frac{x_0xk}{z_0}\Bigr)}I_0}^{Interference}\\
%%\nonumber
%&&-\overbrace{\frac{\alpha z_rAk^2}{2\pi z_0^4}\Bigl(x_0^2+z_0^2+x^2+z^2-2z_0z+2x_0x\Bigr)\cos{\Bigl(k(\frac{r^2-2y_0y-2z_0z}{2z_0}+\frac{x_0x}{z_0})\Bigr)}I_0+\frac{\alpha A^2k^3 }{4\pi^2 z_0^5}\Bigl(x_0^2+z_0^2+x^2+z^2-2z_0z\Bigr)I_0}^{Imbalance},
\end{eqnarray}
\end{widetext}
\normalsize
where we have assumed $I_0=\epsilon_0 c E_0^2/2$, $\alpha=\Delta x x_0 k/z_0$, $(x_0,\hspace{1mm} y_0,\hspace{1mm} x,\hspace{1mm} y,\hspace{1mm} z)  << z_0$, $1/r_1\approx 1/z_0$ and $1/r_2\approx 1/z_0$.  Further, we have assumed that the depolarization of the scattered light is negligible. This is true when the levitated particle is small ($a << \lambda$) compared to the trapping laser's wavelength. It can be seen that there are three main terms in the signal that a balanced detector produces. These are: an interference term consisting of the scattered and unscattered light, a term due to the scattered light alone and a term owing to the imbalance ($\alpha > 0$) between the two arms of a balanced detector. In the ideal scenario, where the two arms of a balanced photodetector are perfectly balanced $\Delta I_{Imb}=0$ ($\alpha=0$). Further, the contribution of the scattering term in the overall signal is much smaller than the interference term. As a result, below we only analyse the interference term and the term due to the imbalance, and show their importance in the context of balanced detection.

Expanding $\cos{\Bigl(k\frac{r^2-2y_0y-2z_0z}{2z_0}\Bigr)}$, $\sin{\Bigl(k\frac{r^2-2y_0y-2z_0z}{2z_0}\Bigr)}$, $\cos{\Bigl(\frac{x_0xk}{z_0}\Bigr)}$ and $\sin{\Bigl(\frac{x_0xk}{z_0}\Bigr)}$ into their respective Taylor's series and keeping only lower order terms, and substituting $x=A_x\sin{\omega_x t}$, $y=A_y\sin{\omega_y t}$ and $z=A_z\sin{\omega_z t}$, the interference term can be written as (see appendix Eq \ref{eqn109} for details)

\tiny
\begin{eqnarray}
\nonumber
\Delta I_{Inter}&=&-\frac{x_0z_rAk^3}{\pi z_0^5}I_0\Bigl[(x_0^2+z_0^2)A_x\sin{\omega_xt}-2z_0A_xA_z\cos{(\omega_x-\omega_z)t}\\
&&+2z_0A_xA_z\cos{(\omega_x+\omega_z)t}+\mathbf{f(\omega_x,\omega_y,\omega_z)}]I_0,
%&\approx&\frac{x_0z_rAk^3}{\pi z_0^5}\Bigl[(x_0^2+z_0^2)A_x\sin{\omega_xt}+2z_0A_xA_z\cos{(\omega_x+\omega_z)t}\\
%&&-2z_0A_xA_z\cos{(\omega_x-\omega_z)t}+\mathbf{f(\omega_x,\omega_y,\omega_z)}\Bigr] I_0,
\label{eqn16a}
\end{eqnarray}
\normalsize
where we have assumed $\phi_x=\phi_y=\phi_z=0$ for simplicity.

\begin{figure}[h]
\subfigure{
\includegraphics[width=8.5cm]{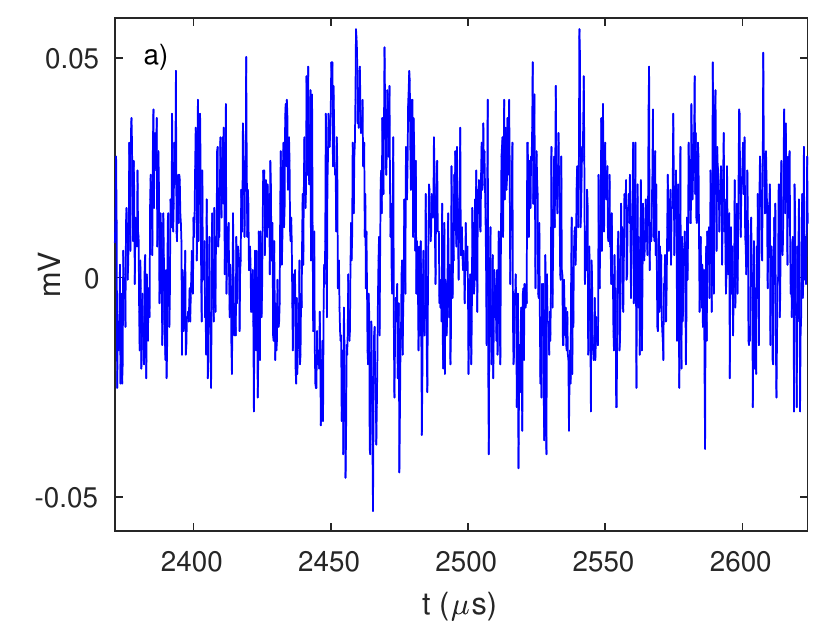}}
\subfigure{\includegraphics[width=8.5cm]{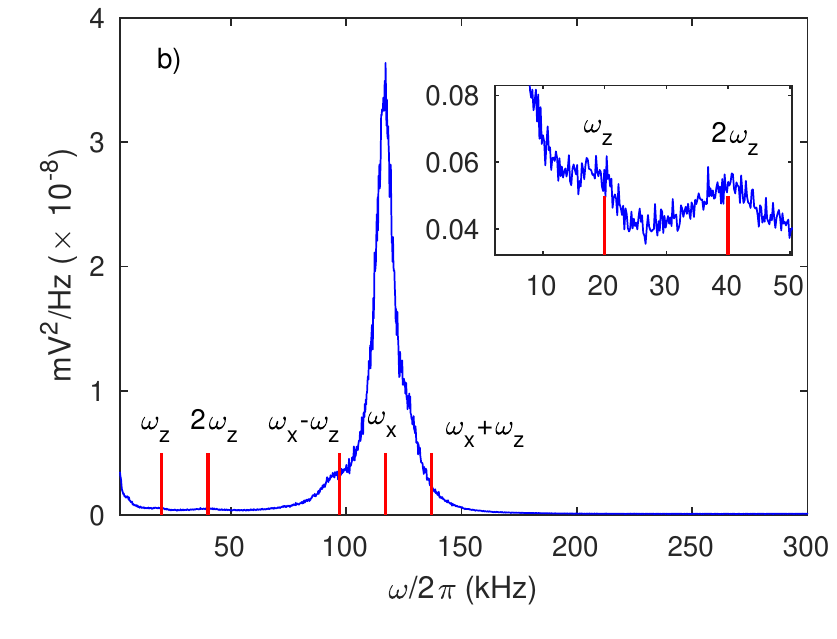}}
\caption{A levitated silica nanoparticle ($50$ nm) at 3 mbar of pressure - a) time trace as it oscillates inside the trap, and b) power spectral density. Red vertical lines in b) represent some of the frequencies (except $\omega_z$ and its harmonics) predicted by Eq. (\ref{eqn16a}). Data were collected at 3 mbar of pressure.}
\label{fig2}
\end{figure}

From Eq. (\ref{eqn16a}), one can find that even though the balanced detector in the configuration shown in Fig. \ref{fig1}, is meant to detect the oscillation frequency along the $x-$axis, our model predicts the detection of many other frequencies $\mathbf{f(\omega_x,\omega_y,\omega_z)}$ in addition to $\omega_x$. To justify the validity of Eq. (\ref{eqn16a}), Fig. \ref{fig2}b shows a Fourier transform of the measured time domain signal obtained using a balanced photodiode (PDB210C/M - Large-Area balanced photodetector, Thorlabs Ltd) from our levitated experiment. In this particular case, a $50$ nm silica particle was levitated using a dipole trap and data were collected at $3$ mbar of pressure. Immediately, one can recognize the desired frequency along the $x$-axis, $\omega_x/2\pi$. One can also find two shoulders at $\omega_x-\omega_z$ and $\omega_x+\omega_z$ as predicted in Eqn. (\ref{eqn16a}). These frequencies are much weaker than $\omega_x$ as understandable from Eq. (\ref{eqn16a}). Elaborately, from our experiment we have $A_x\approx A_y\approx A_z/2\approx 100$ nm, $x_0=y_0=1$ mm, $r_0\approx z_0=10$ mm. On substitution of these values in Eq. (\ref{eqn16a}), one finds the ratio of the amplitudes of $\omega_x-\omega_z$ or $\omega_x+\omega_z$, and $\omega_x$ is $2A_z/z_0\approx 4\times 10^{-5}$ which is small and only in qualitative agreement with our experimental data (see Fig. \ref{fig2}b). Mismatch between the ratios of the amplitudes of the experimental data and the theoretical model can be attributed to the different approximations and assumptions we have made in deriving the theoretical model. Other frequencies as appeared in Eqn. (\ref{eqn16a}) are about two orders of magnitude weaker than $\omega_x-\omega_z$ or $\omega_x+\omega_z$. This is good for parametric feedback cooling where frequencies other than the desired frequency are problematic. A consequence of the unwanted frequencies is that they impart amplitude modulation to the intensity of the desired signal as can be seen in Fig. \ref{fig2}a. This has been observed in earlier experiments\cite{Gieseler2012, GieselerNatPhys2013} as well.

The appearance of $\omega_z$ and its harmonics in Fig. \ref{fig2} are not expected according to Eq. (\ref{eqn16a}). Nevertheless, it can be explained by analysing the impact of the imbalance between the two arms of a balanced detector. Specifically, in theory $\alpha=0$ is achievable but in a realistic laboratory environment a minor imbalance between the two detectors is unavoidable. The consequence of this unwanted imbalance can be quite significant. For example, the ratio between the dominant imbalance ($\omega_z$, see appendix Eq. \ref{eqn110}) and interference ($\omega_x$, $1^{st}$ term in Eq. \ref{eqn16a}) terms is $\approx 2\Delta x/z_0$. If one considers $\Delta x=0.01z_0$ then the ratio of these two terms is $\approx 0.02$. This is equivalent to $2\%$ of the intensity along the $x$-axis and is non-trivial. 

\begin{figure}
\subfigure{
\includegraphics[width=8.5cm]{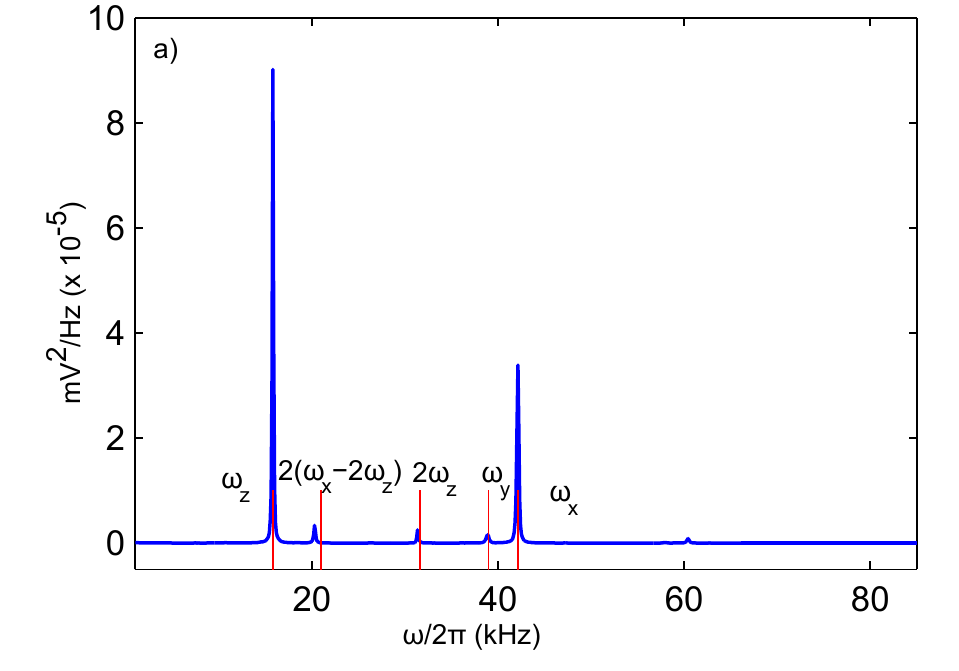}}
\subfigure{
\includegraphics[width=8.50cm]{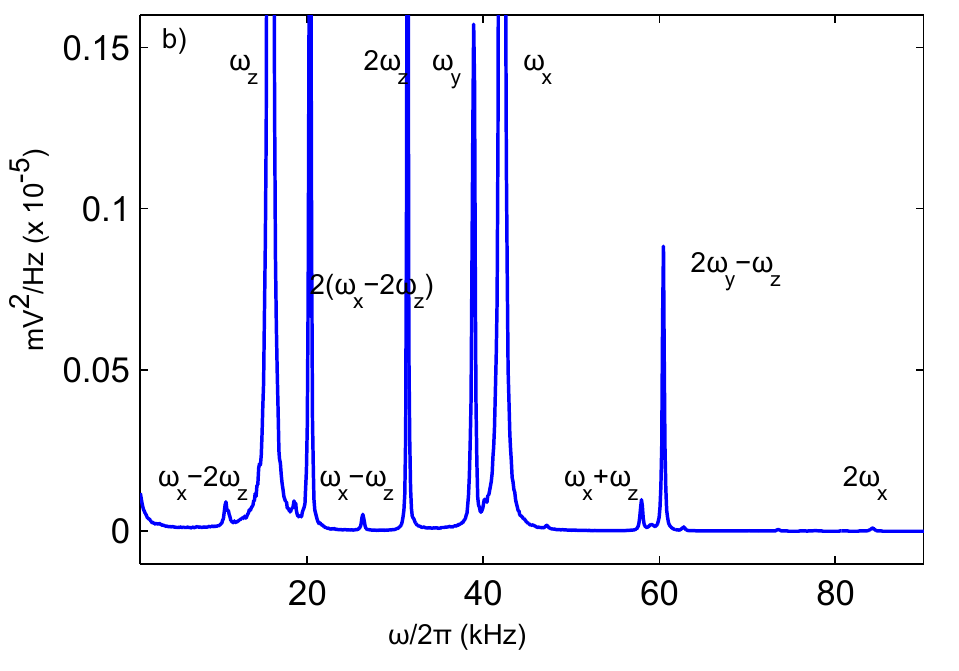}}
\caption{Power spectral density (PSD) of a relatively large (380 nm) silica nanoparticle at 0.50 mbar in a dipole trap - a) shows most of the dominant frequencies visible in the PSD while b) is the zoomed view of a).}
\label{fig3}
\end{figure}

For larger particles spurious frequencies become even more pronounced as we show in Fig. \ref{fig3}. In this example, data were collected using a $380$ nm silica particle and the detector was set to detect the frequency along the $x-$axis. One can see that the intensities of $\omega_z$ and its harmonics as well as other frequencies are comparable to the intensity of $\omega_x$. From our laboratory experience, this happens with the majority of the larger nanoparticles that we levitate using our dipole trap. A similar phenomenon has also been detected by other groups \cite{LiNatPhys2011}. We believe that for the larger particles it is relatively easy to move outside the linear region of the trap to the non-linear part. This introduces coupling between the different modes of oscillations\cite{GieselerNatNano2014,GieselerPRL2014} and hence the appearances of frequencies other than the desired one. It is also plausible that strong scattering from large particles and the ensuing interference around the trapping region alters the trapping potential profile which introduces coupling between different axes that is otherwise assumed decoupled. We also believe that larger particles modify the propagation path of the trapping light due to refraction more strongly than their smaller counterparts. This creates severe dynamic imbalance between the two arms of a detector as the particles oscillate inside the trap and leads to the appearance of unwanted frequencies. Further, as the trapped particle becomes large, the scattered light from the particle gets depolarized\cite{BohrenHuffman}. As a result the interference between the scattered and the trapping light diminishes. Further, the assumption that the electric field remains constant over the distance the particle traverses breaks down.

In the extreme case of imbalance where $\alpha>>0$, the balanced detector shown in Fig. \ref{fig1} turns into an oscillation detector along the $z-$axis. Specifically, in the balanced detection of frequency along the $z-$axis, one arm of the balanced detector is fed with a fixed laser light which does not go through the trap while the other arm of the detector is illuminated with the scattered plus the directly transmitted light that goes through the trap\cite{Gieseler2012}. The role of the constant laser power in the first arm is to cancel the dc term that arises in the second photodiode. The overall model is shown Eq. (\ref{eqn18}) (see appendix Eq. \ref{eqn111} for derivation).

\tiny
\begin{eqnarray}
\nonumber
\Delta I_{z}&\approx&\frac{z_r(x_0^2+z_0^2)k^3A}{4\pi z_0^5}(A_x^2\sin^2{\omega_xt}+A_y^2\sin^2{\omega_yt}+A_z^2\sin^2{\omega_zt}\\
&&-2x_0A_x\sin{\omega_xt}-2y_0A_y\sin{\omega_yt}-2z_0A_z\sin{\omega_zt})I_0.
\label{eqn18}
\end{eqnarray}
\normalsize

\begin{figure}[h]
\includegraphics[width=8.5cm]{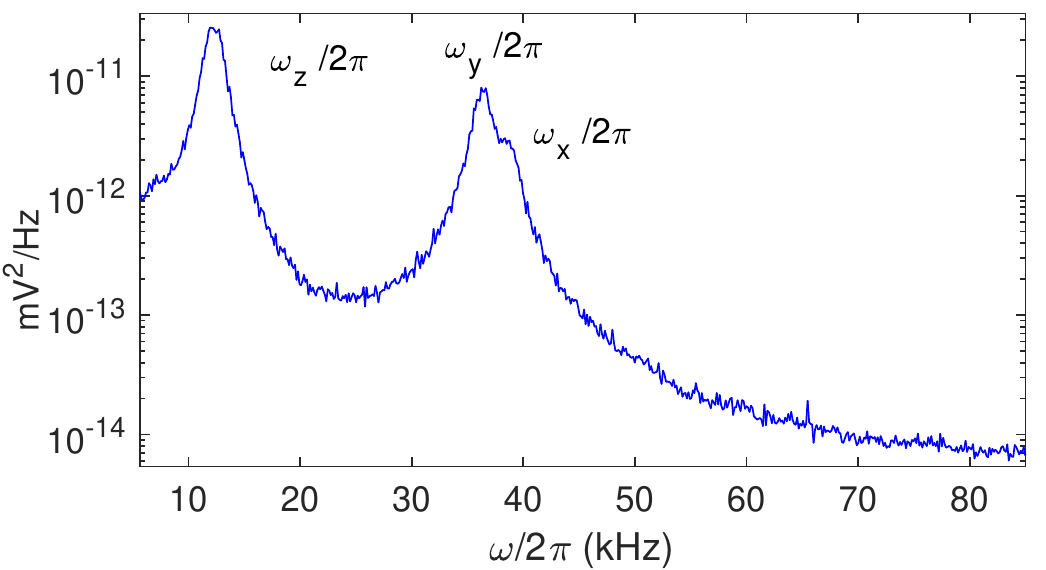}
\caption{Power spectral density from a balanced detector along the $z-$axis.}
\label{fig3a}
\end{figure}

One can see that the detector for the $z-$axis detects the frequency along the desired axis as well as frequencies along the remaining two axes. Our model of the detector also predicts the detection of harmonics of the fundamental modes albeit very weakly. Experimental data from the $z-$ axis detector in our levitated setup is shown in Fig. \ref{fig3a}. In agreement with the model, in our experiment we detect all three frequencies along the three laboratory axes. Our model also agrees with the experimental power spectral density data presented by Li et al. in Ref. 2 where frequencies along all three axes are visible.

Finally, it is instructive to consider the impact of the unwanted frequencies in parametric feedback cooling - particularly in experiments where large particles are levitated. It is well known that as the particle size increases the separation in frequency among the different oscillation axes diminishes. These waning gaps in frequency require a proportional reduction in the bandwidths (BW) of the filters used in parameteric feedback cooling. At some point a further reduction of the BW becomes un-viable and filters become ineffective in suppressing unwanted frequencies. Observing this phenomenon in our levitated experiments we wanted to derive an analytical formulation that can predict the achievable CM temperature under certain frequency noises. However, we find that an analytical model of this situation can only be derived if the spurious frequencies are the harmonics of the fundamental modes ($\omega_x$, $\omega_y$ and $\omega_z$) that one wants to cool. But this is not the case for these experiments. As a result, we are unable to provide an analytical solution of this situation. Nevertheless, we believe that an electrodynamic numerical simulation, which is not considered here, can provide quantitative answers of the impact of the unwanted frequencies in parametric feedback cooling.    

\section{Conclusions}

We have developed a model which represents the combined interferometric and balanced detection schemes used in levitated optomechanics. According to our model, frequencies such as $\omega_x$ and its harmonics as well as the sum and differences of $\omega_x$ with the frequencies of oscillation along the remaining two axes and their harmonics are naturally expected from a balanced detector along the $x-$axis. However, the appearances of $\omega_y$ and $\omega_z$ and their harmonics in the detector along the $x-$axis can be attributed to the imbalance present in the detection system. An effect of these unwanted frequencies is the reduction of the signal to noise ratio which might limit the ultimate temperature achievable in parametric feedback cooling. This is particularly true for systems involving large levitated particles. According to our model designing the experiment to make $x_0k/z_0$ smaller will tend to reduce problems with imbalance as $\alpha=\Delta x x_0 k/z_0$. In particular, making $k$ smaller by using a longer wavelength laser for trapping seems appropriate. Reducing $x_0$ would also help but will reduce the amount of light that a detector receives.  

%\bibliography{C:/Bibliography/MainBib}
%\bibliography{MainBib}

%merlin.mbs aipnum4-1.bst 2010-07-25 4.21a (PWD, AO, DPC) hacked
%Control: key (0)
%Control: author (8) initials jnrlst
%Control: editor formatted (1) identically to author
%Control: production of article title (-1) disabled
%Control: page (0) single
%Control: year (1) truncated
%Control: production of eprint (0) enabled
%

\begin{widetext}
\appendix
\section{}
Exploiting the impulse response of free space propagation\cite{SalehTeich}, the directly transmitted light received by the photodiodes can be expressed as
\tiny
\begin{eqnarray}
\nonumber
E_{T_1}(x_0,y_0,z_0)&=&\frac{i\exp{[i\omega t-ikz_0]}}{\lambda z_0}\int_\infty^\infty\int_\infty^\infty{E_0\exp{[-\frac{x^2+y^2}{w_0^2}]}}\exp{[-i\pi\frac{(x_0-x)^2+(y_0-y)^2}{\lambda z_0}]}dx dy\\
\nonumber
&=&\frac{iE_0\exp{[i\omega t-ikz_0]}}{\lambda z_0} \exp{[-i\pi\frac{x_0^2+y_0^2}{\lambda z_0}]}\int_\infty^\infty\int_\infty^\infty{\exp{[-(\frac{1}{w_0^2}+\frac{i\pi}{\lambda z_0})x^2+i\frac{2\pi x_0}{\lambda z_0}x]}\exp{[-(\frac{1}{w_0^2}+\frac{i\pi}{\lambda z_0})y^2+i\frac{2\pi y_0}{\lambda z_0}y]}}dxdy\\
\nonumber
&=&\frac{iE_0\exp{[i\omega t-ikz_0]}}{\lambda z_0} \exp{[-i\pi\frac{x_0^2+y_0^2}{\lambda z_0}]}\frac{\pi\lambda z_0w_0^2}{\lambda z_0+i\pi w_0^2}\exp{[-\frac{\pi w_0^2(x_0^2+y_0^2)}{\lambda z_0(\lambda z_0+i\pi w_0^2)}]}\\
\nonumber
&=&\frac{i\pi w_0^2}{\lambda z_0+i\pi w_0^2} E_0\exp{[i\omega t-ikz_0]}\exp{[-i\pi\frac{x_0^2+y_0^2}{\lambda z_0}]}\exp{[-\frac{\pi w_0^2(x_0^2+y_0^2)}{\lambda^2 z_0^2}]}\\
\nonumber
&=&\frac{E_0}{1-i\frac{z_0}{z_r}}\exp{\Bigl[i\omega t-ikz_0(1+\frac{x_0^2+y_0^2}{2z_0^2})\Bigr]}\exp{[-\frac{\pi w_0^2(x_0^2+y_0^2)}{\lambda^2 z_0^2}]}\\
\nonumber
&\approx&\frac{E_0}{1-i\frac{z_0}{z_r}}\exp{\Bigl[i\omega t-ikz_0(1+\frac{x_0^2+y_0^2}{2z_0^2})\Bigr]}\\
%\nonumber
%&=&\frac{E_0}{1-i\frac{z_0}{z_r}}\exp{\Bigl[i\omega t-ikz_0(1+\frac{x_0^2+y_0^2}{2z_0^2})\Bigr]}\\
\nonumber
&\approx&\frac{z_rE_0}{z_0}\exp{\Bigl[i\omega t-ikz_0(1+\frac{x_0^2+y_0^2}{2z_0^2})+\arctan{(\frac{z_0}{z_r})}\Bigr]}\\
\nonumber
&\approx&\frac{z_rE_0}{z_0}\exp{\Bigl[i\omega t-ikz_0(1+\frac{x_0^2+y_0^2}{2z_0^2})+\pi/2\Bigr]}\\
%\nonumber
%&\approx&i\frac{z_rE_0}{z_0}\exp{\Bigl[i\omega t-ikz_0(1+\frac{x_0^2+y_0^2}{2z_0^2})\Bigr]}\\
%\nonumber
\Re\Big\{\mathbf{E_{T_1}}\Big\}&=&\frac{z_r}{z_0}E_0\sin{\Bigl(\omega t-kz_0(1+\frac{x_0^2+y_0^2}{2z_0^2})\Bigr)}\mathbf{\hat{y}}
%&\approx&-\frac{z_r}{z_0}E_0\cos{\Bigl(\omega t-kz_0(1+\frac{x_0^2+y_0^2}{2z_0^2})\Bigr)}\mathbf{\hat{y}}
\label{eqn100}
\end{eqnarray}

\normalsize

where we have used $\int^\infty _\infty{\exp{[-ax^2+ibx]} dx}=\sqrt{\pi/a}\exp{[-b^2/(4a)]}$, $\frac{w_0^2}{\lambda r_0} << 1$ and $\exp{[-\frac{\pi w_0^2(x_0^2+y_0^2)}{\lambda^2 r_0^2}]}\approx 1$. Similarly, $E_{T_2}$ can be expressed as

%% Further, in the last line we have assumed\cite{SalehTeich} the Fresnel number $(x_0^2+y_0^2)/(\lambda r_0)<< 1$.

\tiny
\begin{eqnarray}
E_{T_2}(-x_0-\Delta x,y_0,z_0)&\approx&\frac{z_rE_0}{z_0}\exp{\Bigl[i\omega t-ikz_0(1+\frac{x_0^2+y_0^2}{2z_0^2})+\pi/2\Bigr]}\exp{(-i\frac{\Delta x x_0k}{z_0})}
%\nonumber
%&=&i\frac{z_rE_0}{z_0}\exp{\Bigl[i\omega t-ikz_0(1+\frac{x_0^2+y_0^2}{2z_0^2})\Bigr]}\exp{(-i\frac{\Delta x x_0k}{z_0})}\\
%\nonumber
%\Re\Bigl\{\mathbf{E_{T_2}}\Bigr\}&\approx &i\frac{z_r}{z_0}E_0\sin{(\frac{\Delta x x_0k}{z_0})}\sin{\Bigl(\omega t-kz_0(1+\frac{x_0^2+y_0^2}{2z_0^2})\Bigr)}\mathbf{\hat{y}}\\
%&&-\frac{z_r}{z_0}E_0\cos{(\frac{\Delta x x_0k}{z_0})}\cos{\Bigl(\omega t-kz_0(1+\frac{x_0^2+y_0^2}{2z_0^2})\Bigr)}\mathbf{\hat{y}}
\label{eqn101}
\end{eqnarray}
\normalsize
 
Scattered field received by the two photodiodes can be expressed as\cite{BohrenHuffman} 

\tiny
\begin{eqnarray}
\nonumber
\mathbf{E_{s_1}}&=&\frac{A k^2 E_0}{4\pi r_1} \exp{\{i(\omega t-kr_1)\}}\Bigl [\frac{(x_0-x)(y_0-y)}{r_1^2}\mathbf{\hat{x}}-\frac{(x_0-x)^2+(z_0-z)^2}{r_1^2}\mathbf{\hat{y}}+\frac{(z_0-z)(y_0-y)}{r_1^2}\mathbf{\hat{z}}\Bigr]\\
\nonumber
&\approx&-\frac{A k^2\bigl[(x_0-x)^2+(z_0-z)^2\bigr]}{4\pi r_1^3} E_0\exp{\{i(\omega t-kr_1)\}}\mathbf{\hat{y}}\\
\nonumber
&=&-\Bigl[\frac{A(x_0^2+z_0^2)k^2}{4\pi r_1^3}+\frac{A(x^2+z^2-2x_0x-2z_0z)k^2}{4\pi r_1^3}\Bigr]E_0\exp{\{i(\omega t-kz_0(1+\frac{x_0^2+y_0^2}{2z_0^2}+\frac{r^2}{2z_0^2}-\frac{x_0x+y_0y+z_0z}{z_0^2}))\}}\mathbf{\hat{y}}\\
\nonumber
\Re\Bigl\{\mathbf{E_{s_1}}\Bigr\}&=&-\Bigl[\frac{A(x_0^2+z_0^2)k^2}{4\pi r_1^3}+\frac{A(x^2+z^2-2x_0x-2z_0z)k^2}{4\pi r_1^3}\Bigr]E_0\cos{\{(\omega t-kr_1)\}}\mathbf{\hat{y}}\\
\nonumber
&\approx&-\Bigl[\frac{A(x_0^2+z_0^2)k^2}{4\pi r_0^3}-\frac{A(x_0x+z_0z)k^2}{2\pi r_0^3}\Bigr]E_0\cos{\{(\omega t-kr_1)\}}\mathbf{\hat{y}}\\
&\approx&-\Bigl[\frac{A(x_0^2+z_0^2)k^2}{4\pi r_0^3}-\frac{z_0zAk^2}{2\pi r_0^3}-\frac{x_0xAk^2}{2\pi r_0^3}\Bigr]E_0\cos{\{(\omega t-kz_0(1+\frac{x_0^2+y_0^2}{2z_0^2}+\frac{r^2}{2z_0^2}-\frac{x_0x+y_0y+z_0z}{z_0^2}))\}}\mathbf{\hat{y}}
\label{eqn102}
\end{eqnarray}
\normalsize

and 
\tiny
\begin{eqnarray}
\nonumber
\mathbf{E_{s_2}}&=&\frac{A k^2 E_0}{4\pi r_2} \exp{\{i(\omega t-kr_2)\}}\Bigl [-\frac{(x_0+\Delta x+x)(y_0-y)}{r_2^2}\mathbf{\hat{x}}-\frac{(x_0+\Delta x+x)^2+(z_0-z)^2}{r_2^2}\mathbf{\hat{y}}+\frac{(z_0-z)(y_0-y)}{r_2^2}\mathbf{\hat{z}}\Bigr]\\
\nonumber
&\approx&-\frac{A k^2 E_0}{4\pi r_2} \exp{\{i(\omega t-kr_2)\}}\Bigl [\frac{(x_0+\Delta x+x)^2+(z_0-z)^2}{r_2^2}\Bigr]\mathbf{\hat{y}}\\
\nonumber
&=&-\Bigl[\frac{A (x_0^2+z_0^2)k^2 }{4\pi r_2^3}+\frac{A (x^2+z^2+2x_0x-2z_0z)k^2 }{4\pi r_2^3}+\frac{A (\Delta x^2+2x_0\Delta x +2x\Delta x)k^2 }{4\pi r_2^3}\Bigr] E_0\exp{\{i(\omega t-kr_2)\}} \mathbf{\hat{y}}\\
\nonumber
&\approx&-\Bigl[\frac{A (x_0^2+z_0^2)k^2 }{4\pi r_2^3}+\frac{A (x^2+z^2+2x_0x-2z_0z)k^2 }{4\pi r_2^3}+\frac{\Delta xx_0Ak^2 }{2\pi r_0^3}\Bigr] E_0\exp{\{i(\omega t-kr_2)\}} \mathbf{\hat{y}}\\
\nonumber
&=&-\Bigl[\frac{A (x_0^2+z_0^2)k^2 }{4\pi r_2^3}+\frac{A (x^2+z^2-2z_0z)k^2 }{4\pi r_2^3}+\frac{x_0xAk^2 }{2\pi r_2^3}+\frac{\Delta xx_0Ak^2 }{2\pi r_2^3}\Bigr] E_0\exp{\{i(\omega t-kz_0(1+\frac{x_0^2+y_0^2}{2z_0^2}+\frac{r^2}{2z_0^2}+\frac{x_0x-y_0y-z_0z}{z_0^2}+\frac{\Delta x x_0}{z_0^2}))\}} \mathbf{\hat{y}}\\
\Re\Bigl\{\mathbf{E_{s_2}}\Bigl\}&=&-\Bigl[\frac{A (x_0^2+z_0^2)k^2 }{4\pi r_0^3}-\frac{z_0zAk^2 }{2\pi r_2^3}+\frac{x_0xAk^2 }{2\pi r_2^3}+\frac{\Delta xx_0Ak^2 }{2\pi r_2^3}\Bigr] E_0\cos{\{(\omega t-kz_0(1+\frac{x_0^2+y_0^2}{2z_0^2}+\frac{r^2}{2z_0^2}+\frac{x_0x-y_0y-z_0z}{z_0^2}+\frac{\Delta x x_0}{z_0^2}))\}} \mathbf{\hat{y}},
\label{eqn103}
\end{eqnarray}
\normalsize
where $r_1\approx z_0+\frac{x_0^2+y_0^2}{2z_0}+\frac{r^2}{2z_0}-\frac{x_0x+y_0y+z_0z}{z_0}$ and $r_2\approx z_0+\frac{x_0^2+y_0^2}{2z_0}+\frac{r^2}{2z_0}+\frac{x_0x-y_0y-z_0z}{z_0}+\frac{\Delta x x_0}{z_0}$.

Considering the scattered field, and the field due to the directly transmitted light ($\mathbf{E_{T_1}}$ or $\mathbf{E_{T_2}}$) together, the overall field at the two photodiodes can be written as 

\tiny
\begin{eqnarray}
\nonumber
\mathbf{E_{D_1}}&\approx&\frac{z_rE_0}{z_0}\exp{\Bigl[i\omega t-ikz_0(1+\frac{x_0^2+y_0^2}{2z_0^2})+\pi/2\Bigr]}-\Bigl[\frac{A (x_0^2+z_0^2)k^2 }{4\pi r_2^3}+\frac{A (x^2+z^2-2z_0z)k^2 }{4\pi r_2^3}-\frac{x_0xAk^2 }{2\pi r_2^3}\Bigr]\\
\nonumber
&& \times E_0\exp{\{i(\omega t-kz_0(1+\frac{x_0^2+y_0^2}{2z_0^2}+\frac{r^2}{2z_0^2}-\frac{x_0x+y_0y+z_0z}{z_0^2}))\}} \mathbf{\hat{y}}\\
\nonumber
\mathbf{E_{D_2}}&\approx&  \frac{z_rE_0}{z_0}\exp{\Bigl[i\omega t-ikz_0(1+\frac{x_0^2+y_0^2}{2z_0^2})+\pi/2\Bigr]}\exp{(-i\frac{\Delta x x_0k}{z_0})}-\Bigl[\frac{A (x_0^2+z_0^2)k^2 }{4\pi r_2^3}+\frac{A (x^2+z^2-2z_0z)k^2 }{4\pi r_2^3}+\frac{x_0xAk^2 }{2\pi r_2^3}+\frac{\Delta xx_0Ak^2 }{2\pi r_2^3}\Bigr]\\
&& \times E_0\exp{\{i(\omega t-kz_0(1+\frac{x_0^2+y_0^2}{2z_0^2}+\frac{r^2}{2z_0^2}+\frac{x_0x-y_0y-z_0z}{z_0^2}+\frac{\Delta x x_0}{z_0^2}))\}} \mathbf{\hat{y}}
\label{eqn104}
\end{eqnarray}
\normalsize
where we have assumed $x_0,\hspace{1mm} y_0,\hspace{1mm} x,\hspace{1mm} y,\hspace{1mm} z  << z_0$. Further, we have assumed that the depolarization of the scattered light is negligible. This is true when the levitated particle is small ($a << \lambda$) compared to the trapping laser's wavelength. The respective intensities can be expressed as 
\tiny
\begin{eqnarray}
\nonumber
I_{D_1}&=&\frac{\epsilon_0 c E_{D_1}E_{D_1}^{*}}{2}\\ 
\nonumber
&=&\frac{\epsilon_0 cz_r^2E_0^2}{2z_0^2}+\Bigl[\frac{A(x_0^2+z_0^2)k^2}{4\pi z_0^3}+\frac{A(x^2+z^2-2z_0z)k^2}{4\pi z_0^3}-\frac{x_0xAk^2 }{2\pi z_0^3}\Bigr]^2\frac{\epsilon_0 cE_0^2}{2}\\
\nonumber
&&-2\frac{z_r\epsilon_0 cE_0^2}{2z_0}\Bigl[\frac{A(x_0^2+z_0^2)k^2}{4\pi z_0^3}+\frac{A(x^2+z^2-2z_0z)k^2}{4\pi z_0^3}-\frac{x_0xAk^2 }{2\pi z_0^3}\Bigr]\sin{\Bigl(k(\frac{r^2}{2z_0}-\frac{x_0x+y_0y+z_0z}{z_0})\Bigr)}\\
\nonumber
&\approx&\frac{\epsilon_0 cz_r^2E_0^2}{2z_0^2}+\Bigl[\frac{A(x_0^2+z_0^2)k^2}{4\pi z_0^3}+\frac{A(x^2+z^2-2z_0z)k^2}{4\pi z_0^3}-\frac{x_0xAk^2 }{2\pi z_0^3}\Bigr]^2\frac{\epsilon_0 cE_0^2}{2}\\
\nonumber
&&-2\frac{z_r\epsilon_0 cE_0^2}{2z_0}\Bigl[\frac{A(x_0^2+z_0^2)k^2}{4\pi z_0^3}+\frac{A(x^2+z^2-2z_0z)k^2}{4\pi z_0^3}-\frac{x_0xAk^2 }{2\pi z_0^3}\Bigr]\sin{\Bigl(k(\frac{r^2-2y_0y-2z_0z}{2z_0}-\frac{x_0x}{z_0})\Bigr)}\\
\nonumber
&=&\frac{\epsilon_0 cz_r^2E_0^2}{2z_0^2}+\Bigl[\frac{A(x_0^2+z_0^2)k^2}{4\pi z_0^3}+\frac{A(x^2+z^2-2z_0z)k^2}{4\pi z_0^3}-\frac{x_0xAk^2 }{2\pi z_0^3}\Bigr]^2\frac{\epsilon_0 cE_0^2}{2}\\
\nonumber
&&-2\Bigl[\frac{A(x_0^2+z_0^2)k^2}{4\pi z_0^3}+\frac{A(x^2+z^2-2z_0z)k^2}{4\pi z_0^3}\Bigr]\sin{\Bigl(k(\frac{r^2-2y_0y-2z_0z}{2z_0}-\frac{x_0x}{z_0})\Bigr)}\frac{z_r\epsilon_0 cE_0^2}{2z_0}\\
%\nonumber
&&+2\frac{x_0xAk^2 }{2\pi z_0^3}\sin{\Bigl(k(\frac{r^2-2y_0y-2z_0z}{2z_0}-\frac{x_0x}{z_0})\Bigr)}\frac{z_r\epsilon_0 cE_0^2}{2z_0}
\label{eqn106}
\end{eqnarray}
\normalsize
 and
 \tiny
\begin{eqnarray}
\nonumber
I_{D_2}&=&\frac{\epsilon_0 c E_{D_2}E_{D_2}^*}{2}\\
\nonumber
&=&\frac{\epsilon_0 cz_r^2E_0^2}{2z_0^2}+\Bigl(\frac{A (x_0^2+z_0^2)k^2 }{4\pi z_0^3}+\frac{A (x^2+z^2-2z_0z)k^2 }{4\pi z_0^3}+\frac{x_0xAk^2 }{2\pi z_0^3}+\frac{\Delta xx_0Ak^2 }{2\pi z_0^3}\Bigr)^2\frac{\epsilon_0 cE_0^2}{2}\\
\nonumber
&&-2\Bigl(\frac{A (x_0^2+z_0^2)k^2 }{4\pi z_0^3}+\frac{A (x^2+z^2-2z_0z)k^2 }{4\pi z_0^3}+\frac{x_0xAk^2 }{2\pi z_0^3}+\frac{\Delta xx_0Ak^2 }{2\pi z_0^3}\Bigr)\frac{z_r\epsilon_0 c E_0^2}{2z_0}\sin{\Bigl(k(\frac{r^2}{2z_0}+\frac{x_0x-y_0y-z_0z}{z_0}+\frac{\Delta x x_0}{z_0})\Bigr)}\\
\nonumber
&=&\frac{\epsilon_0 cz_r^2E_0^2}{2z_0^2}+\Bigl(\frac{A (x_0^2+z_0^2)k^2 }{4\pi z_0^3}+\frac{A (x^2+z^2-2z_0z)k^2 }{4\pi z_0^3}+\frac{x_0xAk^2 }{2\pi z_0^3}+\frac{\Delta xx_0Ak^2 }{2\pi z_0^3}\Bigr)^2\frac{\epsilon_0 cE_0^2}{2}\\
\nonumber
&&-2\Bigl(\frac{A (x_0^2+z_0^2)k^2 }{4\pi z_0^3}+\frac{A (x^2+z^2-2z_0z)k^2 }{4\pi z_0^3}+\frac{x_0xAk^2 }{2\pi z_0^3}+\frac{\Delta xx_0Ak^2 }{2\pi z_0^3}\Bigr)\frac{z_r\epsilon_0 c E_0^2}{2z_0}\sin{\Bigl(k(\frac{r^2-2y_0y-2z_0z}{2z_0}+\frac{x_0x}{z_0}+\frac{\Delta x x_0}{z_0})\Bigr)}\\
\nonumber
&=&\frac{\epsilon_0 cz_r^2E_0^2}{2z_0^2}+\Bigl(\frac{A (x_0^2+z_0^2)k^2 }{4\pi z_0^3}+\frac{A (x^2+z^2-2z_0z)k^2 }{4\pi z_0^3}+\frac{x_0xAk^2 }{2\pi z_0^3}+\frac{\Delta xx_0Ak^2 }{2\pi z_0^3}\Bigr)^2\frac{\epsilon_0 cE_0^2}{2}\\
\nonumber
&&-2\Bigl(\frac{A (x_0^2+z_0^2)k^2 }{4\pi z_0^3}+\frac{A (x^2+z^2-2z_0z)k^2 }{4\pi z_0^3}+\frac{x_0xAk^2 }{2\pi z_0^3}+\frac{\Delta xx_0Ak^2 }{2\pi z_0^3}\Bigr)\frac{z_r\epsilon_0 c E_0^2}{2z_0}\Bigl[\sin{\Bigl(k(\frac{r^2-2y_0y-2z_0z}{2z_0}+\frac{x_0x}{z_0})\Bigr)}\cos{(\frac{\Delta x x_0k}{z_0})}\\
\nonumber
&&+\cos{\Bigl(k(\frac{r^2-2y_0y-2z_0z}{2z_0}+\frac{x_0x}{z_0})\Bigr)}\sin{(\frac{\Delta x x_0k}{z_0})}\Bigr]\\
\nonumber
&\approx&\frac{\epsilon_0 cz_r^2E_0^2}{2z_0^2}+\Bigl(\frac{A (x_0^2+z_0^2)k^2 }{4\pi z_0^3}+\frac{A (x^2+z^2-2z_0z)k^2 }{4\pi z_0^3}+\frac{x_0xAk^2 }{2\pi z_0^3}+\frac{\Delta xx_0Ak^2 }{2\pi z_0^3}\Bigr)^2\frac{\epsilon_0 cE_0^2}{2}\\
\nonumber
&&-2\Bigl(\frac{A (x_0^2+z_0^2)k^2 }{4\pi z_0^3}+\frac{A (x^2+z^2-2z_0z)k^2 }{4\pi z_0^3}+\frac{x_0xAk^2 }{2\pi z_0^3}+\frac{\Delta xx_0Ak^2 }{2\pi z_0^3}\Bigr)\frac{z_r\epsilon_0 c E_0^2}{2z_0}\Bigl[\sin{\Bigl(k(\frac{r^2-2y_0y-2z_0z}{2z_0}+\frac{x_0x}{z_0})\Bigr)}\\
\nonumber
&&+\frac{\Delta x x_0k}{z_0}\cos{\Bigl(k(\frac{r^2-2y_0y-2z_0z}{2z_0}+\frac{x_0x}{z_0})\Bigr)}\Bigr]\\
\nonumber
&=&\frac{\epsilon_0 cz_r^2E_0^2}{2z_0^2}+\Bigl(\frac{A (x_0^2+z_0^2)k^2 }{4\pi z_0^3}+\frac{A (x^2+z^2-2z_0z)k^2 }{4\pi z_0^3}+\frac{x_0xAk^2 }{2\pi z_0^3}+\frac{\Delta xx_0Ak^2 }{2\pi z_0^3}\Bigr)^2\frac{\epsilon_0 cE_0^2}{2}\\
\nonumber
&&-2\Bigl(\frac{A (x_0^2+z_0^2)k^2 }{4\pi z_0^3}+\frac{A (x^2+z^2-2z_0z)k^2 }{4\pi z_0^3}+\frac{x_0xAk^2 }{2\pi z_0^3}+\frac{\Delta xx_0Ak^2 }{2\pi z_0^3}\Bigr)\sin{\Bigl(k(\frac{r^2-2y_0y-2z_0z}{2z_0}+\frac{x_0x}{z_0})\Bigr)}\frac{z_r\epsilon_0 c E_0^2}{2z_0}\\
\nonumber
&&-\frac{2\Delta x x_0k}{z_0}\Bigl(\frac{A (x_0^2+z_0^2)k^2 }{4\pi z_0^3}+\frac{A (x^2+z^2-2z_0z)k^2 }{4\pi z_0^3}+\frac{x_0xAk^2 }{2\pi z_0^3}+\frac{\Delta xx_0Ak^2 }{2\pi z_0^3}\Bigr)\cos{\Bigl(k(\frac{r^2-2y_0y-2z_0z}{2z_0}+\frac{x_0x}{z_0})\Bigr)}\frac{z_r\epsilon_0 c E_0^2}{2z_0}\\
\nonumber
&=&\frac{\epsilon_0 cz_r^2E_0^2}{2z_0^2}+\Bigl(\frac{A (x_0^2+z_0^2)k^2 }{4\pi z_0^3}+\frac{A (x^2+z^2-2z_0z)k^2 }{4\pi z_0^3}+\frac{x_0xAk^2 }{2\pi z_0^3}+\frac{\Delta xx_0Ak^2 }{2\pi z_0^3}\Bigr)^2\frac{\epsilon_0 cE_0^2}{2}\\
\nonumber
&&-2\Bigl(\frac{A (x_0^2+z_0^2)k^2 }{4\pi z_0^3}+\frac{A (x^2+z^2-2z_0z)k^2 }{4\pi z_0^3}+\frac{x_0xAk^2 }{2\pi z_0^3}\Bigr)\sin{\Bigl(k(\frac{r^2-2y_0y-2z_0z}{2z_0}+\frac{x_0x}{z_0})\Bigr)}\frac{z_r\epsilon_0 c E_0^2}{2z_0}\\
\nonumber
&&-\frac{2\Delta x x_0k}{z_0}\Bigl(\frac{A (x_0^2+z_0^2)k^2 }{4\pi z_0^3}+\frac{A (x^2+z^2-2z_0z)k^2 }{4\pi z_0^3}+\frac{x_0xAk^2 }{2\pi z_0^3}+\frac{\Delta xx_0Ak^2 }{2\pi z_0^3}\Bigr)\cos{\Bigl(k(\frac{r^2-2y_0y-2z_0z}{2z_0}+\frac{x_0x}{z_0})\Bigr)}\frac{z_r\epsilon_0 c E_0^2}{2z_0}\\
\nonumber
&=&\frac{\epsilon_0 cz_r^2E_0^2}{2z_0^2}+\Bigl(\frac{A (x_0^2+z_0^2)k^2 }{4\pi z_0^3}+\frac{A (x^2+z^2-2z_0z)k^2 }{4\pi z_0^3}+\frac{x_0xAk^2 }{2\pi z_0^3}+\frac{\Delta xx_0Ak^2 }{2\pi z_0^3}\Bigr)^2\frac{\epsilon_0 cE_0^2}{2}\\
\nonumber
&&-2\Bigl(\frac{A (x_0^2+z_0^2)k^2 }{4\pi z_0^3}+\frac{A (x^2+z^2-2z_0z)k^2 }{4\pi z_0^3}\Bigr)\sin{\Bigl(k(\frac{r^2-2y_0y-2z_0z}{2z_0}+\frac{x_0x}{z_0})\Bigr)}\frac{z_r\epsilon_0 c E_0^2}{2z_0}\\
\nonumber
&&-2\frac{x_0xAk^2 }{2\pi z_0^3}\sin{\Bigl(k(\frac{r^2-2y_0y-2z_0z}{2z_0}+\frac{x_0x}{z_0})\Bigr)}\frac{z_r\epsilon_0 c E_0^2}{2z_0}\\
%\nonumber
&&-\frac{2\Delta x x_0k}{z_0}\Bigl(\frac{A (x_0^2+z_0^2)k^2 }{4\pi z_0^3}+\frac{A (x^2+z^2-2z_0z)k^2 }{4\pi z_0^3}+\frac{x_0xAk^2 }{2\pi z_0^3}+\frac{\Delta xx_0Ak^2 }{2\pi z_0^3}\Bigr)\cos{\Bigl(k(\frac{r^2-2y_0y-2z_0z}{2z_0}+\frac{x_0x}{z_0})\Bigr)}\frac{z_r\epsilon_0 c E_0^2}{2z_0}.
\label{eqn107}
\end{eqnarray}
\normalsize

\begin{eqnarray}
\nonumber
\Delta I&=& I_{D_2}-I_{D_1}\\
&=&\frac{\epsilon_0 c E_{D_2}E_{D_2}^*}{2}-\frac{\epsilon_0 c E_{D_1}E_{D_1}^*}{2}
\label{eqn107b}
\end{eqnarray}

The difference between $I_{D_2}$ and $I_{D_1}$ is 
\tiny
\begin{eqnarray}
\nonumber
\Delta I&=&\Bigl[\frac{A (x_0^2+z_0^2)k^2 }{4\pi z_0^3}+\frac{A (x^2+z^2-2z_0z)k^2 }{4\pi z_0^3}+\frac{x_0xAk^2 }{2\pi z_0^3}+\frac{\Delta xx_0Ak^2 }{2\pi z_0^3}\Bigr]^2\frac{\epsilon_0 cE_0^2}{2}\\
\nonumber
&&-\Bigl[\frac{A(x_0^2+z_0^2)k^2}{4\pi z_0^3}+\frac{A(x^2+z^2-2z_0z)k^2}{4\pi z_0^3}-\frac{x_0xAk^2 }{2\pi z_0^3}\Bigr]^2\frac{\epsilon_0 cE_0^2}{2}\\
\nonumber
&&-2\Bigl[\frac{A (x_0^2+z_0^2)k^2 }{4\pi z_0^3}+\frac{A (x^2+z^2-2z_0z)k^2 }{4\pi z_0^3}\Bigr]\sin{\Bigl(k(\frac{r^2-2y_0y-2z_0z}{2z_0}+\frac{x_0x}{z_0})\Bigr)}\frac{z_r\epsilon_0 c E_0^2}{2z_0}\\
\nonumber
&&+2\Bigl[\frac{A(x_0^2+z_0^2)k^2}{4\pi z_0^3}+\frac{A(x^2+z^2-2z_0z)k^2}{4\pi z_0^3}\Bigr]\sin{\Bigl(k(\frac{r^2-2y_0y-2z_0z}{2z_0}-\frac{x_0x}{z_0})\Bigr)}\frac{z_r\epsilon_0 cE_0^2}{2z_0}\\
\nonumber
&&-2\frac{x_0xAk^2 }{2\pi z_0^3}\sin{\Bigl(k(\frac{r^2-2y_0y-2z_0z}{2z_0}+\frac{x_0x}{z_0})\Bigr)}\frac{z_r\epsilon_0 c E_0^2}{2z_0}\\
\nonumber
&&-2\frac{x_0xAk^2 }{2\pi z_0^3}\sin{\Bigl(k(\frac{r^2-2y_0y-2z_0z}{2z_0}-\frac{x_0x}{z_0})\Bigr)}\frac{z_r\epsilon_0 cE_0^2}{2z_0}\\
\nonumber
&&-\frac{2\Delta x x_0k}{z_0}\Bigl(\frac{A (x_0^2+z_0^2)k^2 }{4\pi z_0^3}+\frac{A (x^2+z^2-2z_0z)k^2 }{4\pi z_0^3}+\frac{x_0xAk^2 }{2\pi z_0^3}+\frac{\Delta xx_0Ak^2 }{2\pi z_0^3}\Bigr)\cos{\Bigl(k(\frac{r^2-2y_0y-2z_0z}{2z_0}+\frac{x_0x}{z_0})\Bigr)}\frac{z_r\epsilon_0 c E_0^2}{2z_0}\\
\nonumber
&=&\Bigl[\frac{A (x_0^2+z_0^2)k^2 }{4\pi z_0^3}+\frac{A (x^2+z^2-2z_0z)k^2 }{4\pi z_0^3}+\frac{x_0xAk^2 }{2\pi z_0^3}+\frac{\Delta xx_0Ak^2 }{2\pi z_0^3}\Bigr]^2\frac{\epsilon_0 cE_0^2}{2}\\
\nonumber
&&-\Bigl[\frac{A(x_0^2+z_0^2)k^2}{4\pi z_0^3}+\frac{A(x^2+z^2-2z_0z)k^2}{4\pi z_0^3}-\frac{x_0xAk^2 }{2\pi z_0^3}\Bigr]^2\frac{\epsilon_0 cE_0^2}{2}\\
\nonumber
&&-4\Bigl(\frac{A (x_0^2+z_0^2)k^2 }{4\pi z_0^3}+\frac{A (x^2+z^2-2z_0z)k^2 }{4\pi z_0^3}\Bigr)\cos{\Bigl(k\frac{r^2-2y_0y-2z_0z}{2z_0}\Bigr)}\sin{\Bigl(\frac{x_0xk}{z_0}\Bigr)}\frac{z_r\epsilon_0 c E_0^2}{2z_0}\\
\nonumber
&&-4\frac{x_0xAk^2 }{2\pi z_0^3}\sin{\Bigl(k\frac{r^2-2y_0y-2z_0z}{2z_0}\Bigr)}\cos{\Bigl(\frac{x_0xk}{z_0}\Bigr)}\frac{z_r\epsilon_0 c E_0^2}{2z_0}\\
\nonumber
&&-\frac{2\Delta x x_0k}{z_0}\Bigl(\frac{A (x_0^2+z_0^2)k^2 }{4\pi z_0^3}+\frac{A (x^2+z^2-2z_0z)k^2 }{4\pi z_0^3}+\frac{x_0xAk^2 }{2\pi z_0^3}+\frac{\Delta xx_0Ak^2 }{2\pi z_0^3}\Bigr)\cos{\Bigl(k(\frac{r^2-2y_0y-2z_0z}{2z_0}+\frac{x_0x}{z_0})\Bigr)}\frac{z_r\epsilon_0 c E_0^2}{2z_0}\\
\nonumber
&\approx&\overbrace{\frac{x_0xA^2k^4 }{2\pi^2 z_0^6}\Bigl(x_0^2+z_0^2+x^2+z^2-2z_0z\Bigr)I_0}^{Scattering}-\overbrace{\frac{Ak^2}{\pi z_0^3}\Bigl(x_0^2+z_0^2+x^2+z^2-2z_0z\Bigr)\cos{\Bigl(k\frac{r^2-2y_0y-2z_0z}{2z_0}\Bigr)}\sin{\Bigl(\frac{x_0xk}{z_0}\Bigr)}\frac{z_r}{z_0}I_0}^{Interference}\\
\nonumber
&&\overbrace{-4\frac{x_0xAk^2 }{2\pi z_0^3}\sin{\Bigl(k\frac{r^2-2y_0y-2z_0z}{2z_0}\Bigr)}\cos{\Bigl(\frac{x_0xk}{z_0}\Bigr)}\frac{z_r\epsilon_0 c E_0^2}{2z_0}}^{Interference}\\
\nonumber
&&-\overbrace{\frac{\Delta x x_0Ak^3}{2\pi z_0^4}\Bigl(x_0^2+z_0^2+x^2+z^2-2z_0z+2x_0x\Bigr)\cos{\Bigl(k(\frac{r^2-2y_0y-2z_0z}{2z_0}+\frac{x_0x}{z_0})\Bigr)}\frac{z_r}{z_0}I_0}^{Imbalance}\overbrace{\frac{\Delta xx_0A^2k^4 }{4\pi^2 z_0^6}\Bigl(x_0^2+z_0^2+x^2+z^2-2z_0z\Bigr)I_0}^{Imbalance}\\
\nonumber
&=&\overbrace{\frac{x_0xA^2k^4 }{2\pi^2 z_0^6}\Bigl(x_0^2+z_0^2+x^2+z^2-2z_0z\Bigr)I_0}^{Scattering}-\overbrace{\frac{z_rAk^2}{\pi z_0^4}\Bigl(x_0^2+z_0^2+x^2+z^2-2z_0z \Bigr)\Bigr(1 -\frac{k^2(r^2-2y_0y-2z_0z)^2}{8z_0^2}\Bigr)\frac{x_0xk}{z_0} I_0}^{Interference}\\
\nonumber
&&\overbrace{-4\frac{x_0xAk^2 }{2\pi z_0^3}\Bigl(k\frac{r^2-2y_0y-2z_0z}{2z_0}-k^3\frac{(r^2-2y_0y-2z_0z)^3}{48z_0^3}\Bigr)\Bigl(1-\frac{x_0^2x^2k^2}{2z_0^2}\Bigr)\frac{z_r\epsilon_0 c E_0^2}{2z_0}}^{Interference}\\
\nonumber
&&-\overbrace{\frac{\Delta x x_0Ak^3}{2\pi z_0^4}\Bigl(x_0^2+z_0^2+x^2+z^2-2z_0z+2x_0x\Bigr)\cos{\Bigl(k(\frac{r^2-2y_0y-2z_0z}{2z_0}+\frac{x_0x}{z_0})\Bigr)}\frac{z_r}{z_0}I_0}^{Imbalance}+\overbrace{\frac{\Delta xx_0A^2k^4 }{4\pi^2 z_0^6}\Bigl(x_0^2+z_0^2+x^2+z^2-2z_0z\Bigr)I_0}^{Imbalance}\\
\nonumber
&\approx&\overbrace{\frac{x_0xA^2k^4 }{2\pi^2 z_0^6}\Bigl(x_0^2+z_0^2+x^2+z^2-2z_0z\Bigr)I_0}^{Scattering}-\overbrace{\frac{x_0z_rAk^3}{\pi z_0^5}x\Bigl(x_0^2+z_0^2+x^2+z^2-2z_0z\Bigr)\Bigr(1 -\frac{k^2(r^2-2y_0y-2z_0z)^2}{8z_0^2}\Bigr) I_0}^{Interference}\\
\nonumber
&&\overbrace{-4\frac{x_0xAk^2 }{2\pi z_0^3}\Bigl(k\frac{r^2-2y_0y-2z_0z}{2z_0}-k^3\frac{(r^2-2y_0y-2z_0z)^3}{48z_0^3}\Bigr)\Bigl(1-\frac{x_0^2x^2k^2}{2z_0^2}\Bigr)\frac{z_r\epsilon_0 c E_0^2}{2z_0}}^{Interference}\\
%\nonumber
&&-\overbrace{\frac{\Delta x x_0z_rAk^3}{2\pi z_0^5}\Bigl(x_0^2+z_0^2+x^2+z^2-2z_0z+2x_0x\Bigr)\cos{\Bigl(k(\frac{r^2-2y_0y-2z_0z}{2z_0}+\frac{x_0x}{z_0})\Bigr)}I_0}^{Imbalance},
\label{eqn108}
\end{eqnarray}
\normalsize

where $I_0=\epsilon_0 c E_0^2/2$. In order to compare the validity of Eq. \ref{eqn108}, Fig. \ref{fig1b} shows a comparison between Eq. \ref{eqn107b} and Eq \ref{eqn108} where various approximations have been made. It can be seen that they match quite well.

\begin{figure}[h]
	\includegraphics[width=8.5cm]{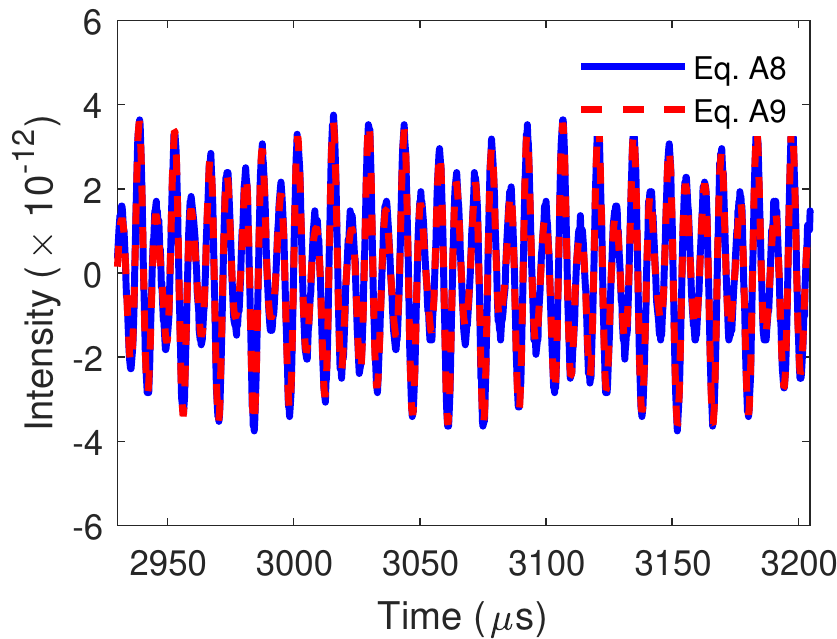}
	\caption{Comparison between the exact difference signal Eq. \ref{eqn107b} (blue solid line) and approximate $\Delta I$ (Eq. \ref{eqn108} ,red broken line) assuming $\alpha=0$, $I_0=2.5\times 10^{11}$ W/m $^2$, $x_0=y_0=1$ mm, $z_0=10$ mm, $A_x=A_y=100$ nm, $A_z=200$ nm, $\omega_x=143$ kHz, $\omega_y=130$ kHz, $\omega_z=33$ kHz and $\phi_x=\phi_y=\phi_z=0$.}
	\label{fig1b}
\end{figure}

From Eq. \ref{eqn108}, the interference term is 
\tiny
\begin{eqnarray}
\nonumber
\Delta I_{Inter}&\approx&-\frac{x_0z_rAk^3}{\pi z_0^5}x\Bigl(x_0^2+z_0^2+x^2+z^2-2z_0z\Bigr)\Bigr(1 -\frac{k^2(r^2-2y_0y-2z_0z)^2}{8z_0^2}\Bigr) I_0\\
\nonumber
&&-2\frac{x_0z_rxAk^2 }{\pi z_0^4}\Bigl(k\frac{r^2-2y_0y-2z_0z}{2z_0}-k^3\frac{(r^2-2y_0y-2z_0z)^3}{48z_0^3}\Bigr)\Bigl(1-\frac{x_0^2x^2k^2}{2z_0^2}\Bigr)I_0\\
\nonumber
&\approx&-\frac{x_0z_rAk^3}{\pi z_0^5}x\Bigl[x_0^2+z_0^2+x^2+z^2-2z_0z -\frac{k^2(x_0^2+z_0^2+x^2+z^2-2z_0z)(r^2-2y_0y-2z_0z)^2}{8z_0^2}\Bigr] I_0\\
\nonumber
&&-2\frac{x_0z_rxAk^2 }{\pi z_0^4}\Bigl(k\frac{r^2-2y_0y-2z_0z}{2z_0}-k^3\frac{(r^2-2y_0y-2z_0z)^3}{48z_0^3}\Bigr)I_0\\
\nonumber
&\approx&-\frac{x_0z_rAk^3}{\pi z_0^5}x\Bigl[x_0^2+z_0^2+x^2+z^2-2z_0z -\frac{k^2(x_0^2+z_0^2)(r^2-2y_0y-2z_0z)^2}{8z_0^2}\Bigr] I_0-\frac{x_0z_rAk^3 }{\pi z_0^5}x(r^2-2y_0y-2z_0z)I_0\\
\nonumber
&=&-\frac{x_0z_rAk^3}{\pi z_0^5}\Bigl[(x_0^2+z_0^2)x+x^3+xz^2-2z_0xz+x(r^2-2y_0y-2z_0z)+\frac{xk^2(x_0^2+z_0^2)(r^2-2y_0y-2z_0z)^2}{8z_0^2}]I_0\\
\nonumber
&=&-\frac{x_0z_rAk^3}{\pi z_0^5}\Bigl[(x_0^2+z_0^2)x+x^3+xz^2-2z_0xz+x^3+xy^2+xz^2-2y_0xy-2z_0xz-\frac{xk^2(x_0^2+z_0^2)(r^2-2y_0y-2z_0z)^2}{8z_0^2}]I_0\\
\nonumber
&=&-\frac{x_0z_rAk^3}{\pi z_0^5}\Bigl[(x_0^2+z_0^2)x-4z_0xz-2y_0xy+2x^3+xy^2+2xz^2-\frac{xk^2(x_0^2+z_0^2)(r^2-2y_0y-2z_0z)^2}{8z_0^2}]I_0\\
\nonumber
&=&-\frac{x_0z_rAk^3}{\pi z_0^5}\Bigl[(x_0^2+z_0^2)A_x\sin{\omega_xt}-4z_0A_xA_z\sin{\omega_xt}\sin{\omega_zt}-2y_0A_xA_y\sin{\omega_xt}\sin{\omega_yt}+2x^3+xy^2+2xz^2-\frac{xk^2(x_0^2+z_0^2)(r^2-2y_0y-2z_0z)^2}{8z_0^2}]I_0\\
\nonumber
&=&-\frac{x_0z_rAk^3}{\pi z_0^5}\Bigl[(x_0^2+z_0^2)A_x\sin{\omega_xt}-4z_0A_xA_z\sin{\omega_xt}\sin{\omega_zt}-2y_0xy+2x^3+xy^2+2xz^2-\frac{xk^2(x_0^2+z_0^2)(r^2-2y_0y-2z_0z)^2}{8z_0^2}]I_0\\
\nonumber
&=&-\frac{x_0z_rAk^3}{\pi z_0^5}\Bigl[(x_0^2+z_0^2)A_x\sin{\omega_xt}-2z_0A_xA_z\cos{(\omega_x-\omega_z)t}+2z_0A_xA_z\cos{(\omega_x+\omega_z)t}-2y_0xy\\
\nonumber
&&+2x^3+xy^2+2xz^2-\frac{xk^2(x_0^2+z_0^2)(r^2-2y_0y-2z_0z)^2}{8z_0^2}]I_0\\
&=&-\frac{x_0z_rAk^3}{\pi z_0^5}\Bigl[(x_0^2+z_0^2)A_x\sin{\omega_xt}-2z_0A_xA_z\cos{(\omega_x-\omega_z)t}+2z_0A_xA_z\cos{(\omega_x+\omega_z)t}+\mathbf{f(\omega_x,\omega_y,\omega_z)}]I_0
\label{eqn109}
\end{eqnarray}
\normalsize

Likewise the term due to the imbalance can be written as 
\tiny
\begin{eqnarray}
\nonumber
\Delta I_{Imb}&=&-\frac{\alpha z_rAk^2}{2\pi z_0^4}\Bigl(x_0^2+z_0^2+x^2+z^2-2z_0z+2x_0x\Bigr)\cos{\Bigl(k(\frac{r^2-2y_0y-2z_0z}{2z_0}+\frac{x_0x}{z_0})\Bigr)}I_0+\frac{\alpha A^2k^3 }{4\pi^2 z_0^5}\Bigl(x_0^2+z_0^2+x^2+z^2-2z_0z\Bigr)I_0\\
\nonumber
&\approx&-\frac{\alpha z_rAk^2}{2\pi z_0^4}\Bigl(x_0^2+z_0^2+x^2+z^2-2z_0z+2x_0x\Bigr)(1-\frac{k^2(r^2-2y_0y-2z_0z+2x_0x)^2}{8z_0^2}\Bigr)I_0\\
\nonumber
&=&-\frac{\alpha z_rAk^2}{2\pi z_0^4}\Bigl[x_0^2+z_0^2+x^2+z^2-2z_0z+2x_0x-\frac{k^2(x_0^2+z_0^2+x^2+z^2-2z_0z+2x_0x)(r^2-2y_0y-2z_0z+2x_0x)^2}{8z_0^2}\Bigr]I_0\\
%\nonumber
&=&-\frac{\alpha z_rAk^2}{2\pi z_0^4}\Bigl[x_0^2+z_0^2-2z_0z+2x_0x+x^2+z^2-\frac{k^2(x_0^2+z_0^2+x^2+z^2-2z_0z+2x_0x)(r^2-2y_0y-2z_0z+2x_0x)^2}{8z_0^2}\Bigr]I_0
\label{eqn110}
\end{eqnarray}
\normalsize

Finally, the signal that a balanced detector along the $z-$axis, excluding the DC component (see main text for details), produces can be expressed as 

\tiny
\begin{eqnarray}
\nonumber
\Delta I_{z}&=&\frac{\epsilon_0 c E_{D_1}E_{D_1}^{*}}{2}\\ 
\nonumber
&=&\frac{\epsilon_0 cz_r^2E_0^2}{2z_0^2}+\Bigl[\frac{A(x_0^2+z_0^2)k^2}{4\pi z_0^3}+\frac{A(x^2+z^2-2z_0z)k^2}{4\pi z_0^3}-\frac{x_0xAk^2 }{2\pi z_0^3}\Bigr]^2\frac{\epsilon_0 cE_0^2}{2}\\
\nonumber
&&+2\frac{z_r\epsilon_0 cE_0^2}{2z_0}\Bigl[\frac{A(x_0^2+z_0^2)k^2}{4\pi z_0^3}+\frac{A(x^2+z^2-2z_0z)k^2}{4\pi z_0^3}-\frac{x_0xAk^2 }{2\pi z_0^3}\Bigr]\sin{\Bigl(k(\frac{r^2}{2z_0}-\frac{x_0x+y_0y+z_0z}{z_0})\Bigr)}\\
\nonumber
&\approx&\frac{\epsilon_0 cz_r^2E_0^2}{2z_0^2}+\Bigl[\frac{A(x_0^2+z_0^2)k^2}{4\pi z_0^3}+\frac{A(x^2+z^2-2z_0z)k^2}{4\pi z_0^3}-\frac{x_0xAk^2 }{2\pi z_0^3}\Bigr]^2\frac{\epsilon_0 cE_0^2}{2}\\
\nonumber
&&+2\frac{z_r\epsilon_0 cE_0^2}{2z_0}\Bigl[\frac{A(x_0^2+z_0^2)k^2}{4\pi z_0^3}+\frac{A(x^2+z^2-2z_0z)k^2}{4\pi z_0^3}-\frac{x_0xAk^2 }{2\pi z_0^3}\Bigr]\sin{\Bigl(k(\frac{r^2-2y_0y-2z_0z}{2z_0}-\frac{x_0x}{z_0})\Bigr)}\\
\nonumber
&=&\frac{\epsilon_0 cz_r^2E_0^2}{2z_0^2}+\Bigl[\frac{A(x_0^2+z_0^2)k^2}{4\pi z_0^3}+\frac{A(x^2+z^2-2z_0z)k^2}{4\pi z_0^3}-\frac{x_0xAk^2 }{2\pi z_0^3}\Bigr]^2\frac{\epsilon_0 cE_0^2}{2}\\
\nonumber
&&+2\Bigl[\frac{A(x_0^2+z_0^2)k^2}{4\pi z_0^3}+\frac{A(x^2+z^2-2z_0z)k^2}{4\pi z_0^3}\Bigr]\sin{\Bigl(k(\frac{r^2-2y_0y-2z_0z}{2z_0}-\frac{x_0x}{z_0})\Bigr)}\frac{z_r\epsilon_0 cE_0^2}{2z_0}\\
\nonumber
&&-2\frac{x_0xAk^2 }{2\pi z_0^3}\sin{\Bigl(k(\frac{r^2-2y_0y-2z_0z}{2z_0}-\frac{x_0x}{z_0})\Bigr)}\frac{z_r\epsilon_0 cE_0^2}{2z_0}\\
\nonumber
&\approx&\Bigl[\frac{A(x_0^2+z_0^2)k^2}{2\pi z_0^3}+\frac{A(x^2+z^2-2z_0z)k^2}{2\pi z_0^3}\Bigr]\sin{\Bigl(\frac{k(r^2-2y_0y-2z_0z-2x_0x)}{2z_0}\Bigr)}\frac{z_r}{z_0}I_0\\
\nonumber
&\approx&\Bigl[\frac{(x_0^2+z_0^2)(r^2-2y_0y-2z_0z-2x_0x)k^3A}{4\pi z_0^4}+\frac{(x^2+z^2-2z_0z)(r^2-2y_0y-2z_0z-2x_0x)k^3A}{4\pi z_0^4}\Bigr]\frac{z_r}{z_0}I_0\\
\nonumber
&\approx&\frac{(x_0^2+z_0^2)(r^2-2y_0y-2z_0z-2x_0x)z_rk^3A}{4\pi z_0^5}I_0\\
&=&\frac{z_r(x_0^2+z_0^2)k^3A}{4\pi z_0^5}(A_x^2\sin^2{\omega_xt}+A_y^2\sin^2{\omega_yt}+A_z^2\sin^2{\omega_zt}-2x_0A_x\sin{\omega_xt}-2y_0A_y\sin{\omega_yt}-2z_0A_z\sin{\omega_zt})I_0
\label{eqn111}
\end{eqnarray}
\end{widetext}
\normalsize

\end{document}